\documentclass[12pt]{iopart}
\usepackage{amsfonts}

\newcommand{\nc}{\newcommand}
\nc{\renc}{\renewcommand}

\nc{\half}{{\textstyle{1\over2}}}
\nc{\ie}{{\it i.e.}}
\nc{\eg}{{\it e.g.}}

\renc{\thefootnote}{\arabic{footnote}}
\nc{\capt}[1]{{\bf Figure.} {\small\sl #1}}

\nc{\eqs}[2]{\mbox{Eqs.~(\ref{#1},\,\ref{#2})}}
\nc{\eq}[1]{\mbox{Eq.~(\ref{#1})}}

\nc{\figs}[2]{\mbox{Figs.~(\ref{#1},\,\ref{#2})}}
\nc{\fig}[1]{\mbox{Fig~.(\ref{#1})}}

\nc{\tag}[1]{\label{#1} \marginpar{{\footnotesize #1}}}
\nc{\mtag}[1]{\label{#1} \mbox{\marginpar{{\footnotesize #1}}}}
\renc{\baselinestretch}{1.5}
\newlength{\overeqskip}
\newlength{\undereqskip}
\nc{\be}[1]{\begin{equation} \mbox{$\label{#1}$}}
\nc{\bea}[1]{\begin{eqnarray} \mbox{$\label{#1}$}}
\nc{\Section}[2]{\section{#2}\label{#1}}
\nc{\Bibitem}[1]{\bibitem{#1}}
\nc{\Label}[1]{\label{#1}}

\nc{\eea}{\vspace{\undereqskip}\end{eqnarray}}
\nc{\ee}{\vspace{\undereqskip}\end{equation}}
\nc{\bdm}{\begin{displaymath}}
\nc{\edm}{\end{displaymath}}
\nc{\dpsty}{\displaystyle}
\nc{\bc}{\begin{center}}
\nc{\ec}{\end{center}}
\nc{\ba}{\begin{array}}
\nc{\ea}{\end{array}}
\nc{\bab}{\begin{abstract}}
\nc{\eab}{\end{abstract}}
\nc{\btab}{\begin{tabular}}
\nc{\etab}{\end{tabular}}
\nc{\bit}{\begin{itemize}}
\nc{\eit}{\end{itemize}}
\nc{\ben}{\begin{enumerate}}
\nc{\een}{\end{enumerate}}
\nc{\bfig}{\begin{figure}}
\nc{\efig}{\end{figure}}
\nc{\arreq}{&\!=\!&}
\nc{\arrmi}{&\!-\!&}
\nc{\arrpl}{&\!+\!&}
\nc{\arrap}{&\!\!\!\approx\!\!\!&}
\nc{\non}{\nonumber\\*}
\nc{\align}{\!\!\!\!\!\!\!\!&&}

\def\lsim{\; \raise0.3ex\hbox{$<$\kern-0.75em
      \raise-1.1ex\hbox{$\sim$}}\; }
\def\gsim{\; \raise0.3ex\hbox{$>$\kern-0.75em
      \raise-1.1ex\hbox{$\sim$}}\; }
\nc{\DOT}{\hspace{-0.08in}{\bf .}\hspace{0.1in}}
\nc{\Laada}{\hbox {$\sqcap$ \kern -1em $\sqcup$}}
\nc\loota{{\scriptstyle\sqcap\kern-0.55em\hbox{$\scriptstyle\sqcup$}}}
\nc\Loota{{\sqcap\kern-0.65em\hbox{$\sqcup$}}}
\nc\laada{\Loota}
\nc{\qed}{\hskip 3em \hbox{\BOX} \vskip 2ex}

\nc{\real}{{\rm I \! R}}
\nc{\Z}{{\sf Z \!\!\! Z}}
\nc{\complex}{{\rm C\!\!\! {\sf I}\,\,}}
\def\bigid{\leavevmode\hbox{\small1\kern-3.8pt\normalsize1}}
\def\id{\leavevmode\hbox{\small1\kern-3.3pt\normalsize1}}
\nc{\slask}{\!\!\!/}
\nc{\bis}{{\prime\prime}}
\nc{\pa}{\partial}
\nc{\na}{\nabla}
\nc{\ra}{\rangle}
\nc{\la}{\langle}
\nc{\goto}{\rightarrow}
\nc{\swap}{\leftrightarrow}

\nc{\EE}[1]{ \mbox{$\cdot10^{#1}$} }
\nc{\abs}[1]{\left|#1\right|}
\nc{\at}[2]{\left.#1\right|_{#2}}
\nc{\norm}[1]{\|#1\|}
\nc{\abscut}[2]{\Abs{#1}_{\scriptscriptstyle#2}}
\nc{\vek}[1]{{\rm\bf #1}}
\nc{\integral}[2]{\int\limits_{#1}^{#2}}
\nc{\inv}[1]{\frac{1}{#1}}
\nc{\dd}[2]{{{\partial #1}\over{\partial #2}}}
\nc{\ddd}[2]{{{{\partial}^2 #1}\over{\partial {#2}^2}}}
\nc{\dddd}[3]{{{{\partial}^2 #1}\over
        {\partial #2 \partial #3}}}
\nc{\dder}[2]{{{d #1}\over{d #2}}}
\nc{\ddder}[2]{{{d^2 #1}\over{d {#2}^2}}}
\nc{\dddder}[3]{{d^2 #1}\over
        {d #2 d #3}}
\nc{\dx}[1]{d\,^{#1}x}
\nc{\dy}[1]{d\,^{#1}y}
\nc{\dz}[1]{d\,^{#1}z}
\nc{\dl}[1]{\frac{d\,^{#1}l}{(2\pi)^{#1}}}
\nc{\dk}[1]{\frac{d\,^{#1}k}{(2\pi)^{#1}}}
\nc{\dq}[1]{\frac{d\,^{#1}q}{(2\pi)^{#1}}}

\nc{\cc}{\mbox{$c.c.$ }}
\nc{\hc}{\mbox{$h.c.$ }}
\nc{\cf}{cf.\ }
\nc{\erfc}{{\rm erfc}}
\nc{\pol}{{\rm pol}}
\nc{\sign}{{\rm sign}}
\nc{\bfT}{{\bf T }}

\nc{\cA}{{\cal A}}
\nc{\cB}{{\cal B}}
\nc{\cD}{{\cal D}}
\nc{\cE}{{\cal E}}
\nc{\cG}{{\cal G}}
\nc{\cH}{{\cal H}}
\nc{\cL}{{\cal L}}
\nc{\cO}{{\cal O}}
\nc{\cT}{{\cal T}}
\nc{\cN}{{\cal N}}
\nc{\rvac}[1]{|{\cal O}#1\rangle}
\nc{\lvac}[1]{\langle{\cal O}#1|}
\nc{\rvacb}[1]{|{\cal O}_\beta #1\rangle}
\nc{\lvacb}[1]{\langle{\cal O}_\beta #1 |}
\nc{\bb}{\bar{\beta}}
\nc{\bt}{\tilde{\beta}}
\nc{\ctH}{\tilde{\cal H}}
\nc{\chH}{\hat{\cal H}}
\nc{\1}{\aa}
\nc{\2}{\"{a}}
\nc{\3}{\"{o}}
\nc{\4}{\AA}
\nc{\5}{\"{A}}
\nc{\6}{\"{O}}
\nc{\al}{\alpha}
\nc{\g}{\gamma}
\nc{\Del}{\Delta}
\nc{\eps}{\epsilon}
\nc{\lam}{\lambda}
\nc{\om}{\omega}
\nc{\Om}{\Omega}
\nc{\ve}{\varepsilon}
\nc{\mn}{{\mu\nu}}
\nc{\ka}{\kappa}
\nc{\vp}{\varphi}

\nc{\advp}[3]{{\it  Adv.\ in\ Phys.\ }{{\bf #1} {(#2)} {#3}}}
\nc{\annp}[3]{{\it  Ann.\ Phys.\ (N.Y.)\ }{{\bf #1} {(#2)} {#3}}}
\nc{\apjl}[3]{{\it  Ap.\ J.\ Lett.\ }{{\bf #1} {(#2)} {#3}}}
\nc{\app}[3]{{\it Astropart.\ Phys.\ }{{\bf #1} {(#2)} {#3}}}
\nc{\cmp}[3]{{\it  Comm.\ Math.\ Phys.\ }{{ \bf #1} {(#2)} {#3}}}
\nc{\cqg}[3]{{\it  Class.\ Quant.\ Grav.\ }{{\bf #1} {(#2)} {#3}}}
\nc{\epl}[3]{{\it  Europhys.\ Lett.\ }{{\bf #1} {(#2)} {#3}}}
\nc{\ijmp}[3]{{\it Int.\ J.\ Mod.\ Phys.\ }{{\bf #1} {(#2)} {#3}}}
\nc{\ijtp}[3]{{\it Int.\ J.\ Theor.\ Phys.\ }{{\bf #1} {(#2)} {#3}}}
\nc{\jmp}[3]{{\it  J.\ Math.\ Phys.\ }{{ \bf #1} {(#2)} {#3}}}
\nc{\jpa}[3]{{\it  J.\ Phys.\ A\ }{{\bf #1} {(#2)} {#3}}}
\nc{\jpc}[3]{{\it  J.\ Phys.\ C\ }{{\bf #1} {(#2)} {#3}}}
\nc{\jap}[3]{{\it J.\ Appl.\ Phys.\ }{{\bf #1} {(#2)} {#3}}}
\nc{\jpsj}[3]{{\it J.\ Phys.\ Soc.\ Japan\ }{{\bf #1} {(#2)} {#3}}}
\nc{\lmp}[3]{{\it Lett.\ Math.\ Phys.\ }{{\bf #1} {(#2)} {#3}}}
\nc{\mpl}[3]{{\it  Mod.\ Phys.\ Lett.\ }{{\bf #1} {(#2)} {#3}}}
\nc{\ncim}[3]{{\it  Nuov.\ Cim.\ }{{\bf #1} {(#2)} {#3}}}
\nc{\np}[3]{{\it  Nucl.\ Phys.\ }{{\bf #1} {(#2)} {#3}}}
\nc{\npps}[3]{{\it  Nucl.\ Phys.\ Proc.\ Suppl.\ }{{\bf #1} {(#2)} {#3}}}
\nc{\pr}[3]{{\it Phys.\ Rev.\ }{{\bf #1} {(#2)} {#3}}}
\nc{\prep}[3]{{\it Phys.\ Rep.\ }{{\bf #1} {(#2)} {#3}}}
\nc{\prsl}[3]{{\it Proc.\ R.\ Soc.\ London\ }{{\bf #1} {(#2)} {#3}}}
\nc{\ptp}[3]{{\it  Prog.\ Theor.\ Phys.\ }{{\bf #1} {(#2)} {#3}}}
\nc{\ptps}[3]{{\it  Prog\ Theor.\ Phys.\ suppl.\ }{{\bf #1} {(#2)} {#3}}}
\nc{\physa}[3]{{\it  Physica\ A\ }{{\bf #1} {(#2)} {#3}}}
\nc{\physb}[3]{{\it  Physica\ B\ }{{\bf #1} {(#2)} {#3}}}
\nc{\phys}[3]{{\it Physica\ }{{\bf #1} {(#2)} {#3}}}
\nc{\rpp}[3]{{\it Rep.\ Prog.\ Phys.\ }{{\bf #1} {(#2)} {#3}}}
\nc{\sjnp}[3]{{\it Sov.\ J.\ Nucl.\ Phys.\ }{{\bf #1} {(#2)} {#3}}}
\nc{\spjetp}[3]{{\it Sov.\ Phys.\ JETP\ }{{\bf #1} {(#2)} {#3}}}
\nc{\yf}[3]{{\it Yad.\ Fiz.\ }{{\bf #1} {(#2)} {#3}}}
\nc{\zetp}[3]{{\it Zh.\ Eksp.\ Teor.\ Fiz.\  }{{\bf #1}  {(#2)} {#3}}}
\nc{\zp}[3]{{\it Z.\ Phys.\ }{{\bf #1} {(#2)} {#3}}}
\nc{\ibid}[3]{{\sl ibid.\ }{{\bf #1} {#2} {#3}}}
\nc{\rf}[1]{(\ref{#1})}
\nc{\nn}{\nonumber \\*}
\nc{\bfB}{\bf{B}}
\nc{\bfv}{\bf{v}}
\nc{\bfx}{\bf{x}}
\nc{\bfy}{\bf{y}}
\nc{\vx}{\vec{x}}
\nc{\vy}{\vec{y}}
\nc{\oB}{\overline{B}}
\nc{\oI}{\overline{I}}
\nc{\oR}{\overline{R}}
\nc{\rar}{\rightarrow}
\nc{\ti}{\times}
\nc{\slsh}{\hskip-5pt/}
\nc{\sm}{Standard~Model~}
\nc{\MP}{M_{\rm Pl}}
\nc{\tp}{t_{\rm Pl}}
\nc{\ave}{\bar{E}}

\nc{\eff}{{\rm eff}}
\nc{\kk}{\vek{k}}
\nc{\pp}{{\rm p}}
\nc{\ga}{g_{a\gamma}}
\nc{\vv}{\\}
\nc{\eee}{{\bf E}}
\nc{\bbb}{{\bf B}}
\nc{\qcd}{T_{\rm QCD}}
\nc{\G}{\rm \ G}
\def\vec#1{{\bf #1}}
\def\vv{\vskip-2pt}

\def\ell{e^{c}LL}

\begin{document}
\hspace{12cm} {\small HIP-2003-59/TH}

\title{Dynamics of MSSM flat directions consisting of multiple scalar fields}

\author{Kari Enqvist}
\address{Physics Department and Helsinki Institute of Physics, P.O.Box 64,
FIN-00014 University of Helsinki, FINLAND}

\author{Asko Jokinen}
\address{Helsinki Institute of Physics, P.O. Box 64, FIN-00014 University
of Helsinki, FINLAND}

\author{Anupam Mazumdar}
\address{Physics Department, McGill University, 3600-University Road,
Montr\'eal, Qu\'ebec, H3A 2T8, CANADA}

\eads{kari.enqvist@helsinki.fi, asko.jokinen@helsinki.fi,
  anupamm@hep.physics.mcgill.ca}

\begin{abstract}
Although often chosen because of simplicity, a single scalar field does not
provide a general parametrization of an MSSM flat direction. We derive a
formalism for a class of gauge invariant polynomials which result in a
multifield description of the flat directions. In contrast to the single field
case, the vanishing of the gauge currents yields an important dynamical
constraint in the multifield framework. We consider in detail the example of
the $H_uL$ flat direction and study the dynamical evolution during and after
inflation. We highlight the differences between the single and the multifield
flat directions. We show that in the multifield case the field space has an
intrinsic curvature and hence unsuppressed non-minimal kinetic terms for the
flat direction scalars arise. Also the phases of the individual components
evolve non-trivially right after inflation, charging the components of the
condensate and producing an enhanced entropy after the decay of the
condensate, which is due to cross-coupling of different lepton flavours in the
F term. However, the qualitative features of the single field Affleck-Dine
baryogenesis, such as the  produced total charge, remain largely unchanged.

\end{abstract}

%

\section{Introduction}

Supersymmetric gauge theories come with gauge invariant polynomials
along which the scalar potential vanishes at a classical level
\cite{Buccella:1981ib}. These correspond to the flat directions in
the moduli space of the scalar fields.  Within the Minimal
Supersymmetric Standard Model (MSSM) the classical potential is given
by the sum of the $F$ and $D$ term contributions, which vanish
individually along these flat directions.  Flatness is however lifted
by soft supersymmetry breaking, but a non-renormalization theorem
guarantees that the flat direction does not obtain any renormalizable
perturbative superpotential correction
\cite{gwr79}. However, non-perturbative (supergravity-induced)
corrections will also lift the flatness.

MSSM flat directions can give rise to a host of cosmologically
interesting dynamics (for a review, see \cite{Enqvist:2003gh}). These
include Affleck-Dine baryogenesis
\cite{Affleck:1984fy,Dine:1995uk,Dine:1995kz}, the cosmological
formation and fragmentation of the MSSM flat direction condensate and
subsequent $Q$-ball formation
\cite{Kusenko:1997zq,Enqvist:1997si,Jokinen:2002xw,Kasuya:1999wu},
reheating the Universe with $Q$-ball evaporation~\cite{Enqvist:2002rj},
generation of baryon isocurvature density perturbations
\cite{Enqvist:1998pf}, as well as curvaton scenarios where MSSM flat directions
reheat the Universe and generate adiabatic density perturbations
\cite{Enqvist:2002rf}. Adiabatic density perturbations
induced by fluctuating inflaton-MSSM flat
direction coupling  has also been discussed~\cite{Enqvist:2003uk}.

In the MSSM all the flat directions have been parameterized by gauge
invariant monomials \cite{Dine:1995kz,Gherghetta:1995dv}. In such a case the
MSSM flat direction is some linear combination of the MSSM scalars and can be
thought of as a trajectory in the moduli space described by a single
scalar degree of freedom. A simple example is the $H_uL$ flat direction $\phi$,
defined as $(H_u)_1 = L_2 =\phi/\sqrt{2}$ (indices here label SU(2)). However,
a flat direction can also be constructed as a gauge invariant polynomial, where
the previously mentioned monomials generate the polynomial. This
means that the flat direction is a non-linear combination of the MSSM
fields. It no longer can be described by a single scalar field. Such a gauge
invariant polynomial, rather than a monomial, is needed e.g. when  for $H_uL$
instead of a single leptonic generation one considers simultaneously all three.

The purpose of the present paper is to illustrate the dynamics of Affleck-Dine
condensate described by a gauge invariant polynomial. Previously Senami and
Yamamoto \cite{Senami:2001qn} have considered such a case in MSSM and its
extensions but they used an extrinsic point of view without gauge invariant
polynomials. We will discuss the dynamics from an intrinsic point of view {\it
  i.e.} using coordinates that describe only the behaviour of the flat
direction. Given that the flat direction is the minimum of energy the two
approaches are equivalent.

We will also discuss how a non-renormalizable superpotential correction
can lift such a multi-field flat direction. We will highlight the formation
and charge conservation of the condensate along with the gauge conditions
required to be satisfied. In the present paper we will mostly restrict
ourselves to a simple example of a flat direction involving $L$ and $H_{u}$
("the $H_{u}L$ flat direction"), which is a linear combination of superfields,
and study its dynamical evolution through the cosmic history of the Universe.

We begin our discussion with the basic definitions of the flat directions in
$N=1$ supersymmetry in Section 2. In Section 3 we first construct $H_{u}L$
flat direction parameterized by a gauge invariant monomial. We then discuss
gauge invariant polynomials which are $F$ and $D$ flat at the level of
unbroken supersymmetry and write down the equation of motion for the $H_{u}L$
scalars with the help of two projection operators, defined in the
Appendix. Vanishing of the gauge currents is carefully accounted for. to In
Section 4 we write down the multifield scalar potential, accounting for
supersymmetry breaking and non-renormalizable superpotential corrections. In
Section 5 we discuss qualitatively the dynamical behavior of the $H_{u}L$
condensate consisting, comparing it with single field case. Section 6 presents
the results numerical studies for the multifield trajectories, charge
densities and charge asymmetries. In Section 7 we make concluding remarks and
discuss the physical implications of the results.

\section{General features of flat directions}

The scalar potential of supersymmetric theories is given by
\cite{Nilles:1983ge}
\be{scapot}
V = \frac{1}{2} \sum_A D^A D^A + \sum_i \left|\dd{W}{\phi_i}\right|^2\,,
\ee
where $W$ is the
superpotential and the D-flat directions are solutions to
\be{dflat21}
D^A = \sum_i \phi_i^{\dagger} T^A \phi_i = 0\,,
\ee
where $\phi_i$ are the scalar fields and $T^A$ are the
generators of the symmetry group, which for $SU(n)$can be chosen
to be Hermitian and traceless. Making a gauge transformation
\be{gaugetr22}
\phi_i \to U(x)\phi_i\,
\ee
transforms the D-term in \eq{dflat21} to
\be{dflat23}
D^A \to \sum_i \phi_i^{\dagger} U(x)^{-1}T^A U(x) \phi_i =
\sum_i \phi_i^{\dagger} \tilde{T}^A \phi_i\,,
\ee
where $\tilde{T}^A$ is also hermitian and traceless. Since $T^A$ are basis
for traceless hermitian matrices, then $\tilde{T}^A$ can be written  as
$\tilde{T}^A= c^{AB}T^B$ with $c^{AB}$ as complex coefficients. Hence the
D-term, \eq{dflat23} reads
\be{dflat24}
D^A \to c^{AB} D^B = 0\,,
\ee
where \eq{dflat21} was used in the last step, proving that
\eq{dflat21} is gauge invariant. The solutions, \eq{dflat24}, are gauge
orbits. (This can be generalized to other gauge groups by virtue of
Baker-Campbell-Hausdorff formula or in general quoting that D term transforms
as a vector in the adjoint representation, which essentially have been shown
to be the case for $SU(N)$).

One should also specify the gauge field configuration related to the scalar
field because it is changed by a gauge transformation. The minimum energy is
obtained with pure gauge configurations $F_{\mu\nu}^A=0$. The equation of
motion for the gauge field is
\be{eqm28}
D^{\nu} F_{\mu\nu}^A = J_{\mu}^A\,,
\ee
where $J_{\mu}^A$ is the matter current defined by
\be{curr29}
J_{\mu}^A = i\sum_i \left((D_{\mu}\phi_i)^{\dagger} T^A \phi_i -
\phi_i^{\dagger} T^A (D_{\mu} \phi_i)\right)\,.
\ee
With a pure gauge configuration the left hand side of \eq{eqm28}
vanishes, providing a constraint on the matter current~\eq{curr29},
and a relationship between the flat direction and gauge fields.
Choosing a gauge corresponds to picking a scalar field value and the
gauge field value from the gauge orbit.

Minimizing the energy of the configurations, \ie the equation of motion,
requires that $D_{\mu}\phi_i=0$. This of course results into vanishing
current in \eq{curr29}. With the gauge choice, $A_{\mu}^A=0$, the scalar
fields would then have to be constants. However, once SUSY breaking is
included, the scalar fields become dynamical and the condition on the minimum
energy becomes more subtle. For a homogeneous scalar field one can check that
the minimum of energy is still obtained by a pure gauge configuration, despite
the fact that the scalar fields have a non-vanishing energy. Therefore the
current~\eq{curr29} has to vanish, but now $D_0 \phi_i\neq 0$. This is a
constraint on the cosmological dynamics of MSSM condensates that has to be
satisfied.

Let us remark that for inhomogeneous fields the situation is even more
complicated. The minimum of energy still obeys the equation of
motion \eq{eqm28}, but it is no longer obvious that the minimum is obtained
with a pure gauge configuration and a zero current. In fact, a more likely
situation is that a spatially varying condensate, implying a spatially varying
gauge charge, will give rise to non-vanishing currents and non-trivial gauge
configurations. In the present paper we do not consider inhomogeneous
condensates, but if one were to study the spatial fluctuations of the flat
direction inhomogeneities would have to be taken into account.

The solutions of~\eq{dflat21} can be parameterized by  analytic gauge invariant
po\-ly\-no\-mials $I(\{\phi_i\})$~\cite{Buccella:1981ib,Sartori:1981zj} which
can be obtained by solving
\be{pol210}
\dd{I}{\phi_i} = C\phi_i^*\,,
\ee
where $C$ is a complex coefficient \footnote{This can be generalized to
the supergravity case with a non-minimal K\"ahler potential, $K$, by
demanding $\partial I/\partial\phi_i = C \partial K/\partial\phi_i$.}.
This equation is also gauge invariant and therefore acts as a solution
to a gauge orbit. The geometrical significance of~\eq{pol210} is that
the polynomial $I$ defines level surfaces, e.g. $I(\phi)=B=const.$,
which actually are the gauge orbits of the complexified gauge group.
Then the flat direction corresponds to the minimum of the norm,
$|\phi|=\sqrt{\sum_i |\phi_i|^2}$, which turns out to be orthogonal
to $\phi$. For analytic functions the complex conjugate of the
gradient is orthogonal to the level surface. This leads to~\eq{pol210} which
just states that the gradient of $I$ and the complex conjugate of $\phi$ have
to be parallel (for details see~\cite{Sartori:1981zj}).

If the polynomial describing the flat direction is actually a monomial and is
composed of $N$ scalar fields of the model, then from~\eq{pol210} it follows
that there are $N$ complex equations for $N+1$ complex variables, $\phi_i$,
and $C$. Therefore there is at least one complex degree of freedom left, which
can be chosen to be the scalar field parameterizing the flat direction. In
case of flat directions parameterized by multiple scalar fields, the relevant
polynomial is a sum of monomials $I_i$ such that $I=\sum_i c_i I_i$, where
$c_i$ are complex coefficients~\footnote{Actually one of the $c_i$ can be
chosen freely, by a reparameterization of $C$ in~\eq{pol210}.}. As a
consequence, there appear extra complex degrees of freedom.


\section{Example: $H_uL$ flat directions}

\subsection{$H_uL$ parameterized by single scalar field}

A generic treatment of polynomials generating MSSM flat directions is beyond
the scope of the present paper. Rather, in order to gain some insight on the
dynamics of flat directions with multiple fields, we focus on a simple
example, the $H_uL$ flat direction. Let us begin by reiterating the case for a
single generation of leptons, denoted here as $L$. The relevant
monomial is then given by~\cite{Dine:1995kz,Gherghetta:1995dv}
\be{mon31}
I = H_uL = \eps_{\al\beta} H_u^{\al} L^{\beta}\,.
\ee
Applying the constraint~\eq{pol210}, one obtains two equations
\bea{constr32}
\dd{I}{H_u^{\al}} &=& \eps_{\al\beta} L^{\beta} = C (H_u^{\al})^*\,, \nn
\dd{I}{L^{\al}} &=& \eps_{\beta\al} H_u^{\beta} = C (L^{\al})^*\,.
\eea
By taking the absolute values of the equations, it follows that
$|C|=1$ and  $|H_u^1|=|L^2|,\,|H_u^2|=|L^1|$. By multiplying the first
equation by $L^{\al}$ and summing over $\al$, one obtains that $H_u\perp L$.
Hence the general structure of $H_uL$ is given by
\be{struct33}
H_u = \left( \ba{ll} \phi \\ \chi \ea \right), \qquad L = e^{i\zeta}
\left( \ba{ll} -\chi^* \\ \phi^* \ea \right)\,,
\ee
where $\phi,\,\chi$ are complex scalar fields and $\zeta$ is a real scalar
field.

Let us now apply the current constraint $J_{\mu}^A=0$ as given by
\eq{curr29}. In the case of single leptonic generation we can write it as
\be{curr34}
\fl J_{\mu}^A = i\left((D_{\mu}H_u)^{\dagger} T^A H_u - H_u^{\dagger} T^A
D_{\mu}H_u + (D_{\mu}L)^{\dagger} T^A L - L^{\dagger}
T^A D_{\mu}L \right) = 0\,,
\ee
where $T^A$ are the $SU(2)\times U(1)$-generators so that there are four
equations altogether. Since the current is gauge invariant, we can choose a
gauge, and since the gauge is pure, we may choose $A_{\mu}^A=0$. (The
details of the calculation is left to the next subsection, see~\eq{struct321}).
As is well known, the final result turns out to be
\bea{struct35}
H_u = \frac{1}{\sqrt{2}} \left( \ba{ll} \phi \\ 0 \ea \right), \qquad L =
\frac{1}{\sqrt{2}} \left( \ba{ll} 0 \\ \phi \ea \right)\,,
\eea
where $\phi$ is a complex scalar field.

One has to check that~\eq{struct35} is also F-flat. The F-terms are
obtained from the superpotential
\be{superp36}
W = \lam_u Q H_u \bar{u} + \lam_d Q H_d \bar{d} + \lam_e L H_d \bar{e} +
\mu_H H_u H_d\,,
\ee
where $\lam_u,\,\lam_d,\,\lam_e$ are the Yukawa couplings, family,
colour and $SU(2)$ indices are suppressed. One can easily find that
the $H_uL$ is automatically F-flat except for the $\mu_H$-term in the
last equation. However, since $\mu_H$ is of the order of the SUSY
breaking mass, and, since SUSY breaking terms anyway lift the flat
direction, the $\mu_H$-terms can be neglected at this stage. The
Lagrangian along the $H_uL$ direction reads simply as
\be{lagr38}
{\cal L} = |\partial_{\mu}H_u|^2 + |\partial_{\mu}L|^2 - V =
|\partial_{\mu}\phi|^2 - V(\phi)\,.
\ee
%

\subsection{$H_uL$ parametrized by three scalar fields}

In principle, the discussion above applies separately for each lepton
generation. However, apart from pure chance, there is no physical mechanism
that would pick out one generation over the others. Hence a most natural
possibility would be to consider $H_uL$ parametrized by all three leptonic
degrees of freedom. Therefore we should consider the gauge invariant polynomial
\be{pol39}
I = \nu_1 H_u L_1 + \nu_2 H_u L_2 + \nu_3 H_u L_3\,,
\ee
where $\nu_i$ are complex coefficients (of which one can be freely chosen).
The D-flatness equations~\eq{pol210} become (compare with \eq{constr32})
\bea{constr310}
\dd{I}{H_u^{\al}} &=& \sum_{i=1}^3 \nu_i \eps_{\al\beta} L_i^{\beta} =
C(H_u^{\al})^*\,, \nn
\dd{I}{L_i^{\al}} &=& \nu_i \eps_{\beta\al} H_u^{\beta} = C(L_i^{\al})^*\,.
\eea
By solving $L_i^{\al}$ from the second equation and inserting into the first,
one obtains the constraint
\be{constr311}
|C|^2 = \sum_{i=1}^3 |\nu_i|^2\,.
\ee
Working the other way around,~\ie~ by solving $H_u^{\al}$ from the first
equation and inserting into the second, one obtains a matrix equation
\be{eq312}
P (\mathbf{L}^{\al})^* = (\mathbf{L}^{\al})^*, \qquad P_{ij} = \frac{\nu_i
  \nu_j^*}{\sum_k |\nu_k|^2}\,,
\ee
where $\mathbf{L}^{\al}=(L_1^{\al},L_2^{\al},L_3^{\al})^T$ is a vector in
flavor space and $P$ is a projection matrix to the vector $\mathbf{\nu} =
(\nu_1,\nu_2,\nu_3)^T$. Note that~\eq{eq312} can be fulfilled only if
$(\mathbf{L}^{\al})^*$ is parallel to $\mathbf{\nu}$ so that
\be{slep313}
(\mathbf{L}^{\al})^* = c^{\al} \mathbf{\nu}\,,
\ee
where $c^{\al}$ is a complex coefficient. The flat direction is given by
\be{flat314}
L_i = \nu_i^* \left( \ba{ll} c^{1*} \\ c^{2*} \ea \right), \quad H_u =
e^{i\delta_C} \sqrt{\sum_i |\nu_i|^2} \left( \ba{ll} c^2 \\ -c^1 \ea
\right)\,,
\ee
where $\delta_C$ is the phase of $C$ which was left undetermined in
\eq{constr311}. Making redefinitions $\phi_i=\nu_i^*c^{2*}$,
$\psi=c^{1*}/c^{2*}$ and $\delta_C = \chi - \delta_2$, where $\chi$ is real
field and $\delta_2$ is the phase of $c^2$, one obtains a simplified
expression for the field content of the $H_uL$-flat direction
\be{flat315}
L_i = \phi_i \left( \ba{ll} \psi \\ 1 \ea \right), \quad H_u = e^{i\chi}
\sqrt{\sum_i |\phi_i|^2} \left( \ba{ll} 1 \\ -\psi^* \ea \right)\,.
\ee

Let us now apply the current constraint $J_{\mu}^A=0$~\eq{curr29}, where
\be{curr316}
\fl J_{\mu}^A = i((D_{\mu}H_u)^{\dagger} T^A H_u - H_u^{\dagger} T^A D_{\mu}
H_u) + i\sum_i ((D_{\mu}L_i)^{\dagger} T^A L_i - L_i^{\dagger} T^A D_{\mu}
L_i)\,.
\ee
We find
\bea{curr317}
\fl J_{\mu}^Y &=& \frac{i}{2} \left[(1+|\psi|^2)\left(\sum_i J_i^\phi - 2i
    \sum_k |\phi_k|^2 \partial_{\mu} \chi\right) + 2\sum_k |\phi_k|^2
  J^\psi\right] = 0\,, \\
\fl J_{\mu}^3 &=& \frac{i}{2} \left[(1-|\psi|^2)\left(\sum_i J_i^\phi - 2i
    \sum_k |\phi_k|^2 \partial_{\mu} \chi\right) - 2\sum_k |\phi_k|^2
  J^\psi\right] = 0\,, \label{curr318} \\
\fl J_{\mu}^{1+i2} &=& i\left[-\psi^*\left(\sum_i J_i^\phi  - 2i \sum_k
    |\phi_k|^2 \partial_{\mu} \chi \right) + 2\sum_k |\phi_k|^2 \partial_{\mu}
    \psi^*\right] =0\,, \label{curr319}
\eea
where
\be{currents}
J_i^\phi \equiv \phi_i^* \partial_{\mu} \phi_i - \phi_i \partial_{\mu}
\phi_i^*,~~J^\psi\equiv \psi^*\partial_{\mu} \psi - \psi \partial_{\mu}
\psi^*\,.
\ee
Adding Eqs.~(\ref{curr317},~\ref{curr318}) together, we obtain
\be{constr320}
\partial_{\mu} \chi = \frac{\sum_j J_j^\phi}{2i \sum_k |\phi_k|^2}\,.
\ee
Using Eqs.~(\ref{constr320},~\ref{curr319}), one obtains
$\partial_{\mu}\psi=0$, so that $\psi=const.$ Therefore we may gauge it away
by a global $U(1)$ transformation including it in the normalization of
$\phi_i$. Hence we finally obtain
\be{struct321}
L_i = \phi_i \left( \ba{ll} 0 \\ 1 \ea \right), \quad H_u = e^{i\chi}
\sqrt{\sum_i |\phi_i|^2} \left( \ba{ll} 1 \\ 0 \ea \right)\,,
\ee
where $\chi$ is given in~\eq{constr320}. Since the final configuration is
three complex dimensional which is the maximal dimension of the flat direction
\cite{Gherghetta:1995dv} containing $H_u$ and $L_i$, the polynomial \eq{pol39}
gives a complete characterization of the $H_uL$ flat direction.

\subsection{The effective Lagrangian for the multiple flat direction}

For one scalar field the constraint for the phase $\chi$ in \eq{constr320}
reduces to $\partial_{\mu}\chi=\partial_{\mu}\theta$, where $\theta$ is the
phase of the scalar field $\phi=|\phi|\exp(i\theta)$. Therefore,  up to a
normalization, the field configuration reduces to \eq{struct35}. However, for
more than one scalar field, the constraint \eq{constr320} is not necessarily
solvable. The problem is the following: on the right hand side there is a
vector field (formed as a sum of the scalar fields $\phi_i$ and their
derivatives), on the left hand side there is a gradient of the scalar
function, so the right hand side has to be rotationless (in the language of
differential forms the one-form on the right-hand side has to be closed for it
to be exact). The necessary (and in simply connected space sufficient)
condition for the vector field to be a gradient of a scalar function is
\be{grad322}
(\partial_{\mu}\partial_{\nu} - \partial_{\nu}\partial_{\mu})\chi = 0\,,
\ee
which yields
\bea{grad323}
\partial_{\mu} \Phi^{\dagger}\, (1-P_1-P_2) \,\partial_{\nu} \Psi = 0\,,
\eea
where the $P_i$ are projection operators discussed in the Appendix.

From \eq{grad323} one sees that the components of the gradient are constrained
to be orthogonal outside of the two-dimensional surface spanned by $\Phi$ and
$\Psi$. On the surface there are no constraints. If there is only one scalar
field, then $\Phi$ and $\Psi$ are two-dimensional and \eq{grad323} is
trivially satisfied, since in this case $P_1+P_2=1$. If the scalar fields do
not depend on space or time, then the constraint is trivially satisfied. If
the fields depend only on one coordinate (time or one of the spatial
coordinates; homogeneous fields fall into this category), then again one can
show that the constraint is trivially satisfied. Since in the cosmological
context the large scale homogeneity is a reasonable approximation, we do not
have to worry here about not fulfilling \eq{grad323}.

The dynamical evolution of the multiple field flat direction is determined by
the MSSM Lagrangian
\be{lagr324}
\mathcal{L} = |\partial_{\mu} H_u|^2 + \sum_{i=1}^3|\partial_{\mu} L_i|^2 -V\,,
\ee
It turns out to be convenient to
change the parameterization of the flat direction by a phase
shift to
\be{struct325}
L_i = e^{-i\chi/2} \phi_i \left( \ba{ll} 0 \\ 1 \ea \right), \quad H_u =
e^{i\chi/2} \sqrt{\sum_i |\phi_i|^2} \left( \ba{ll} 1 \\ 0 \ea \right)\,,
\ee
where $\chi$ is solved from \eq{constr320}. Now the Lagrangian is
obtained by inserting \eq{struct325} into \eq{lagr324}
\be{lagr326}
\mathcal{L} = \frac{1}{2} \partial_{\mu} \Phi^{\dagger} \left(1 + P_1 -
  \frac{1}{2}P_2 \right) \partial^{\mu} \Phi - V\,,
\ee
(for the projection operators $P_i$, see Appendix). Note that there are now
non-minimal kinetic terms. This is due to the fact that the $H_uL$ direction
is a curved sub-manifold of the whole field manifold, as can be seen from
\eq{struct325} which implies that,  up to a gauge choice,  the sub-manifold is
actually a sphere. In the one-field case the flat direction is formed only
along a one-dimensional sub-manifold and therefore has vanishing
curvature. The equations of motion resulting from \eq{lagr326} are (see
Appendix for details)
\bea{eqm327}
\fl \partial_{\mu}\partial^{\mu}\Phi + 3H\dot\Phi +
\left(1-\frac{1}{2}P_1+P_2\right) \dd{V}{\Phi^{\dagger}}
- N^{-1} \left[\partial_{\mu}\Psi\,(\Psi^{\dagger}
  \partial^{\mu}\Phi)\right. \nn \lo+ \left.
  \Psi\,(\partial_{\mu}\Psi^{\dagger} P_2 \partial^{\mu}\Phi) +
  \frac{1}{2}\Phi\,\partial_{\mu} \Phi^{\dagger}
  \left(1-P_1-\frac{3}{2}P_2\right) \partial^{\mu}\Phi \right] = 0,
\eea
where $N=\Phi^{\dagger}\Phi$.

Since we are interested in the background dynamics, the partial derivatives
can be replaced by time derivatives. In that case the different terms have
analogues in classical mechanics. On the first row the potential gradient is
deformed by two projections: $P_1$ makes the potential flatter and $P_2$
steeper in the directions $\Phi$ and $\Psi$ respectively. The last term on the
second row generalizes centripetal acceleration. The rest two terms are
analogous to the Coriolis and Euler forces.

\section{The potential for $H_uL$}

In the real world supersymmetry is broken. This lifts the degeneracy
of the vacuum solutions of the supersymmetric gauge theory, and the
flat directions become dynamical fields
\cite{Dine:1995uk,Dine:1995kz}. Here we just list the contributions
relevant for $H_uL$.

The soft SUSY breaking mass
parameters together with the $\mu$-term read
\be{mass41}
(m_{H_u}^2 + \mu_H^2)|H_u|^2 + \sum_i m_{L_i}^2 |L_i|^2 = \sum_i
m_{\phi_i}^2|\phi_i|^2\,,
\ee
where
\be{mass42}
m_{\phi_i}^2 = m_{H_u}^2 + \mu_H^2 + m_{L_i}^2\,.
\ee
The mass parameters $m_{\phi_i}$ are of the order of the soft SUSY breaking
mass $m_S$, which in the gravity mediated SUSY breaking scenario is the
gravitino mass $m_{3/2}$. The non-renormalizable operators lifting $H_uL$ are
given by \cite{Gherghetta:1995dv},
\be{superp42}
W = \frac{1}{4M} \sum_{ij} \lam_{ij}\, (H_uL_i)\, (H_uL_j)\,,
\ee
where $\lam_{ij}=\lam_{ji}$ are (complex) coupling constants and $M$ is a
large mass scale (typically GUT or Planck scale). The superpotential
\eq{superp42} gives rise to potential terms through the F-terms
\bea{fterm43}
\left|\dd{W}{H_u}\right|^2 + \sum_i \left|\dd{W}{L_i}\right|^2 = \nn
\frac{1}{4M^2} \left[\sum_k |\phi_k|^2 \left|\sum_{ij}\lam_{ij} \phi_i
    \phi_j\right|^2 + \left(\sum_k |\phi_k|^2\right)^2 \sum_i \left|\sum_j
    \lam_{ij} \phi_j\right|^2 \right]\,,
\eea
as well as through the generalized A-terms
\be{aterm44}
Am_{3/2}W + h.c. = \frac{Am_{3/2}}{4M} \sum_k |\phi_k|^2 \sum_{ij} \lam_{ij}
\phi_i \phi_j + h.c.\,,
\ee
where $A\sim 1$ is a complex coefficient. If the K\"ahler potential contains a
non-trivial coupling to the inflaton, such as
\be{kahler45}
K \sim \frac{1}{M_p^2}\,I^{\dagger}I \phi^{\dagger}\phi\,,
\ee
where $I$ is the inflaton and $\phi$ is  $H_u$ or $L_i$, then there is a
mass correction
\be{mass46}
\sum_i c_i H^2 |\phi_i|^2\,,
\ee
where $H$ is the Hubble parameter and $|c_i|\sim 1$ after inflation. During
F-term inflation $|c_i|\sim 1$, but for D-term inflation $|c_i|\ll 1$
\cite{Kolda:1998kc}. Usually we take $c_i=-1$. There are also Hubble induced
A-terms if the inflation is induced by the F-term
\be{aterm47}
aHW + h.c.\,,
\ee
where $a\sim 1$ is a complex coefficient, for D-term inflation $|a|\ll 1$
\cite{Kolda:1998kc}.

\section{Dynamics of the multifield $H_uL$ flat direction}

\subsection{Motion of $H_uL$ during inflation}

During inflation $H\approx H_I = const.$ and the minima of the potential are
fixed points of the equation of motion for the homogeneous mode \eq{eqm327} as
in \cite{Dine:1995kz}. This follows from the fact that $\dot\Phi = \ddot\Phi =
0$ is a fixed point so that eventually only the potential term in
\eq{eqm327} remains in the equation of motion. Local stability is due to the
fact that the last three terms of \eq{eqm327} do not contribute to linearized
perturbations.

\subsection{Motion of $H_uL$ after inflation}

After inflation the Universe is dominated by the inflaton oscillations, which
produces effectively a matter dominated Universe with $H=2/(3t)$. Before the
oscillations start, $H\gg m_S$, we can make the approximation
\cite{Dine:1995kz} that only the Hubble induced mass term \eq{mass46} and the
non-renormalizable terms \eq{fterm43} are important. (The Hubble induced A-term
\eq{aterm47} affects  only the phase). Then we can find a fixed point solution
following \cite{Dine:1995kz} and making the change of
variables
\be{change51}
\phi_i = (MH)^{1/2}\,\psi_i, \qquad t = m_S^{-1}\,e^z\,.
\ee
At the fixed point, $d\chi_i/dz=0$, so that the equation of motion
simplifies to
\be{fixed52}
M^2\left(1 - \frac{1}{2}P_1 + P_2\right) \dd{V_{NR}(\psi)}{\psi^\dagger} -
\left(\frac{4C}{9} + \frac{1}{4}\right)\psi = 0\,,
\ee
where $V_{NR}$ is the potential in \eq{fterm43} with $\psi_i$ as independent
variables instead of $\phi_i$, and $C$ is a matrix with
$C_{ij}=c_i\delta_{ij}$. An effective potential can be obtained for
\eq{fixed52}, whose extrema are fixed points of the equation of motion.
This is a tedious calculation in general but with $c_i=c$ for all $i$
one finds
\be{eff53}
V_{eff} = M^2 V_{NR}(\psi) - 2\left(\frac{4c}{9} + \frac{1}{4}\right)
\psi^{\dagger}\psi\,.
\ee
Hence the effective potential is bounded from below provided
$V_{NR}$ is. The minima of $V_{eff}$ correspond to marginal
fixed points of the equation of motion. This reasoning is similar to the
one field case \cite{Dine:1995kz}, where there is no friction term for
$d=4$ flat directions. However, one cannot solve \eq{eff53} in a closed
form, but solution(s) definitely exist since $V_{NR}$ is positive
definite and higher order than the second term.  When all the terms in the
potential are taken into account, this is only an approximate behavior.

\subsection{The onset of rotation $H\sim m_S$}

At $H\sim m_S$ the scalar fields $\phi_i$ feel the torque produced by the A
terms and start to rotate. The precise onset of rotation depends
on $m_{\phi_i}$ and $c_i$ because the phase rotation of $\phi_i$ starts when
$|c_i|H^2 \sim m_{\phi_i}^2$. When $H\ll m_S$ the lepton charge densities in
the co-moving volume of $\phi_i$ asymptote to (different) constant values. The
scalar fields at this time decay as $a(t)^{-3/2}$ for a matter dominated
Universe. This behavior is similar to the one-field case
\cite{Dine:1995kz,Jokinen:2002xw}. However, for quantitative results one has
to resort to numerics.


\section{Numerical results}

Since we expect a fixed point behavior, for
numerical calculations it is useful to choose the variables as defined in
\eq{change51}. In the one-field case it is easy to obtain the initial
conditions for the evolution after inflation by solving for the field value at
the minimum. Here this is not possible. Instead we assign random initial
values and solve the equations of motion during inflation numerically for 50
e-foldings and then insert the values thus obtained into the equations of
motion valid after inflation. In practice the duration of inflation is
inessential as the fields tend to relax into the minima in 10-20 e-folds.

\begin{figure}
\leavevmode
\centering
\vspace{6cm}
\includegraphics{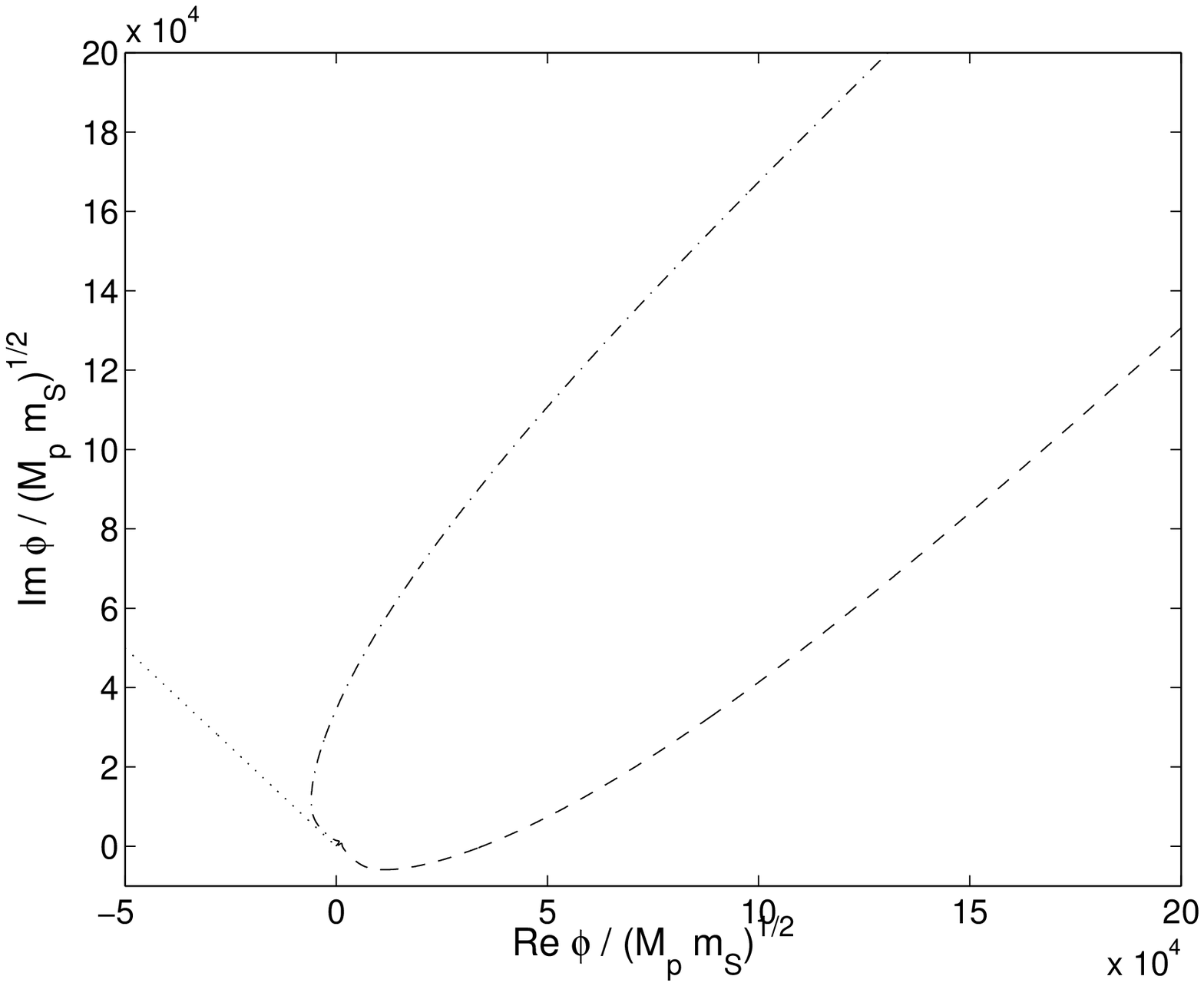}
\includegraphics{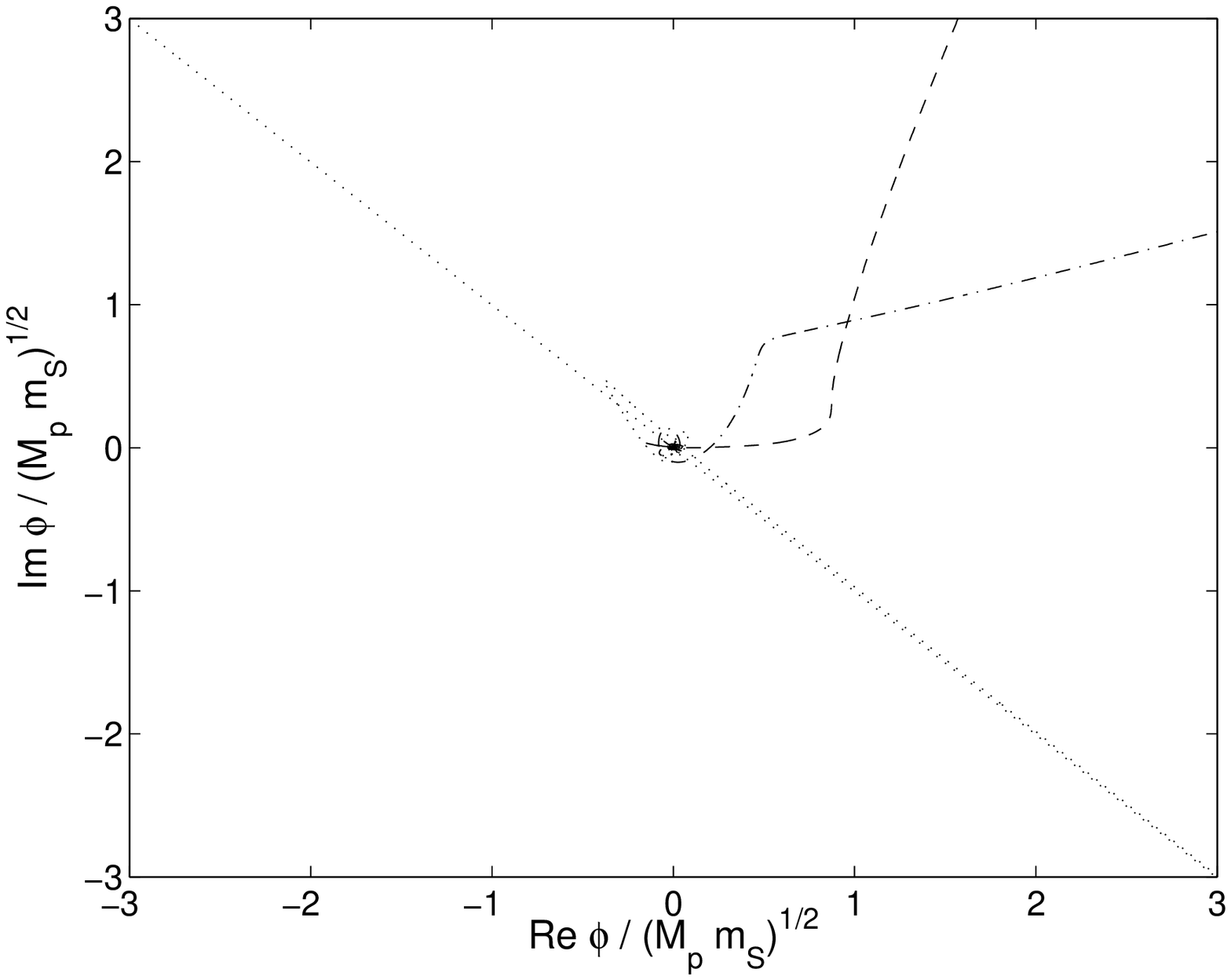}
\caption{\label{fig1} The trajectories of $\phi_i$ for $a=i$, $c_i=A=-1$ and
  $m_{\phi_i}=100\rm{GeV}$ with dashed line for the first, dash-dotted for the
  second and dotted for the third family.}
\end{figure}

\begin{figure}
\leavevmode
\centering
\vspace{12cm}
\includegraphics{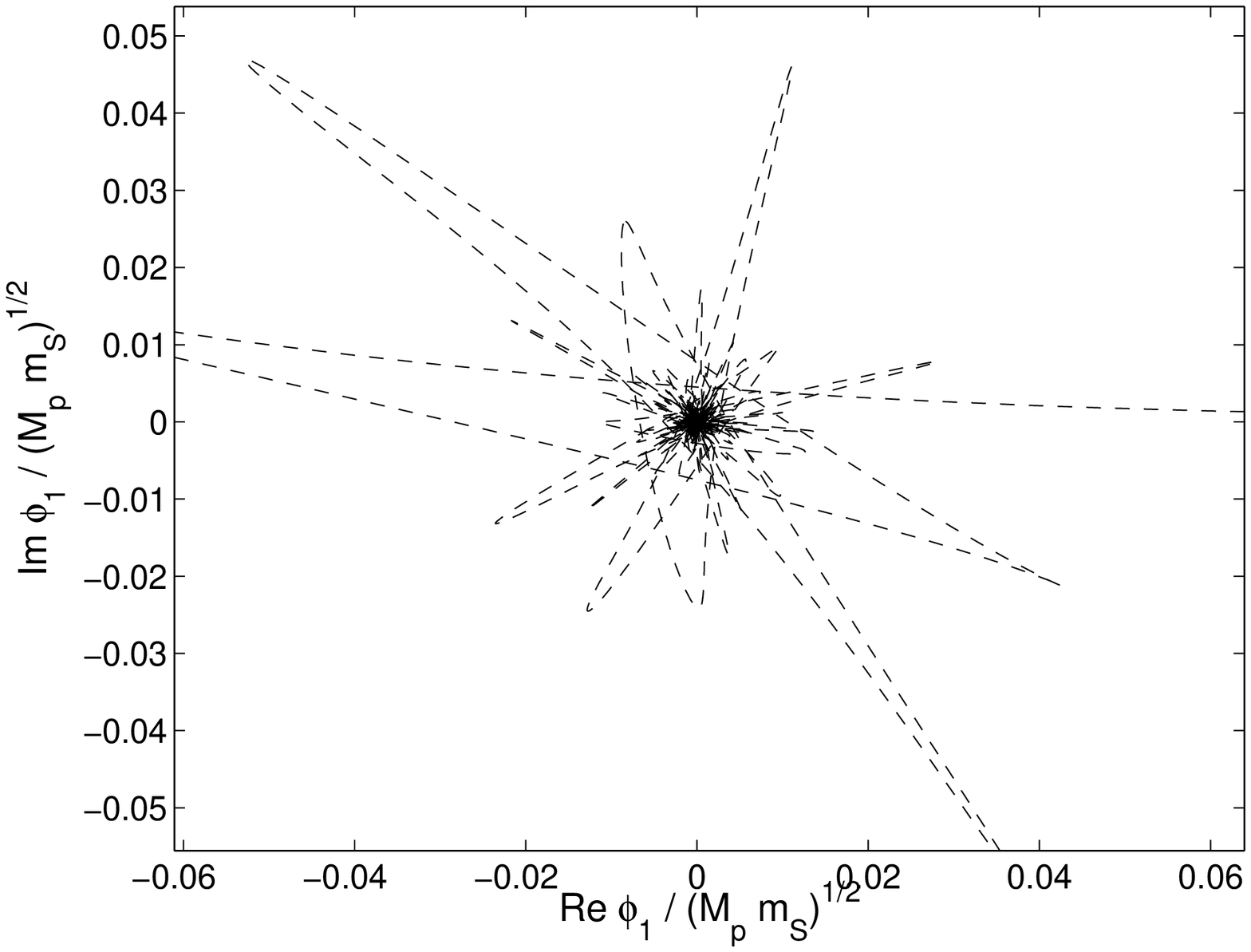}
\includegraphics{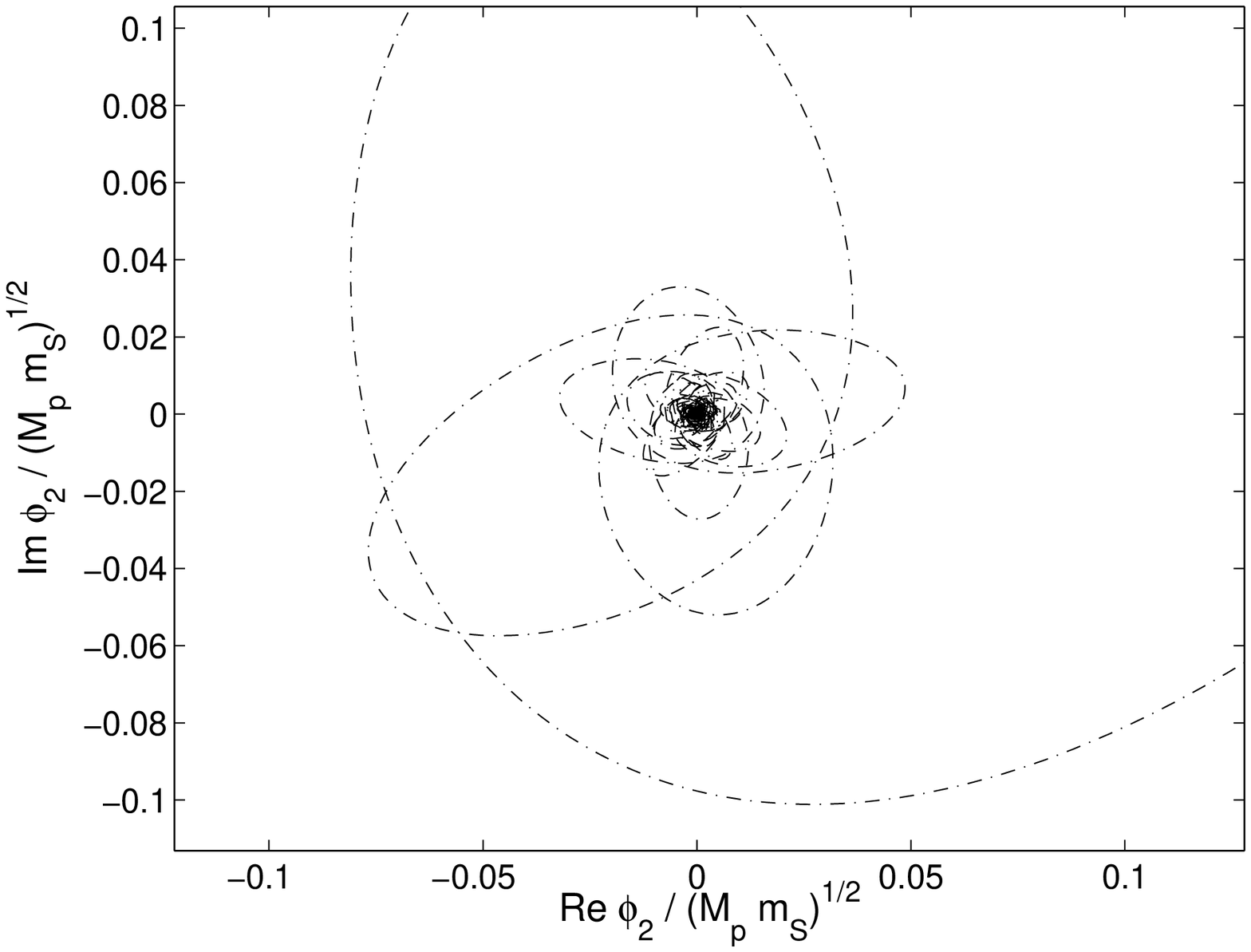}
\includegraphics{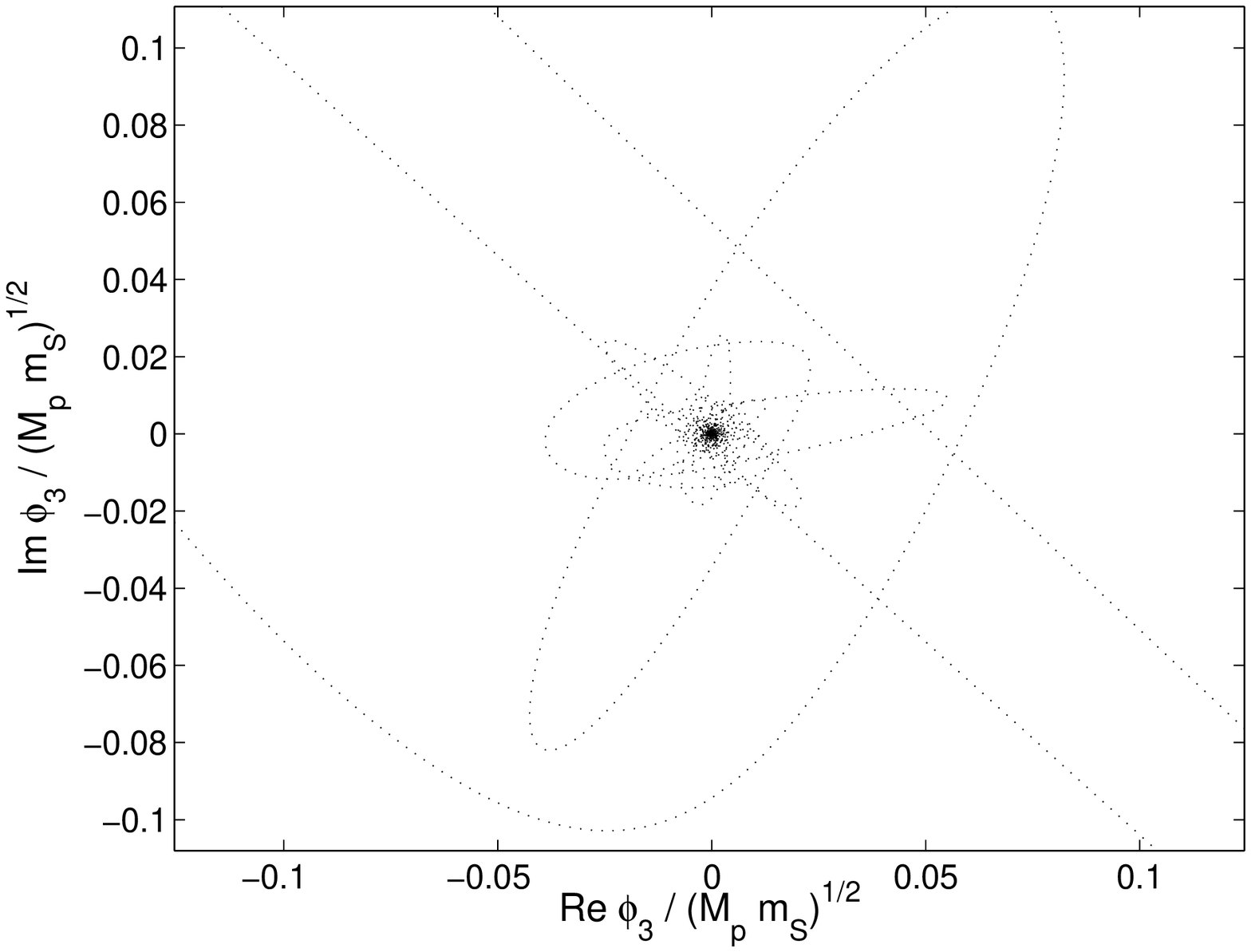}
\caption{\label{fig2} The end trajectories of $\phi_i$ for $a=i$, $c_i=A=-1$
  and $m_{\phi_i}=100\rm{GeV}$ with dashed line for the first, dash-dotted for
  the second and dotted for the third family.}
\end{figure}

\begin{figure}
\leavevmode
\centering
\vspace*{6cm}
\includegraphics{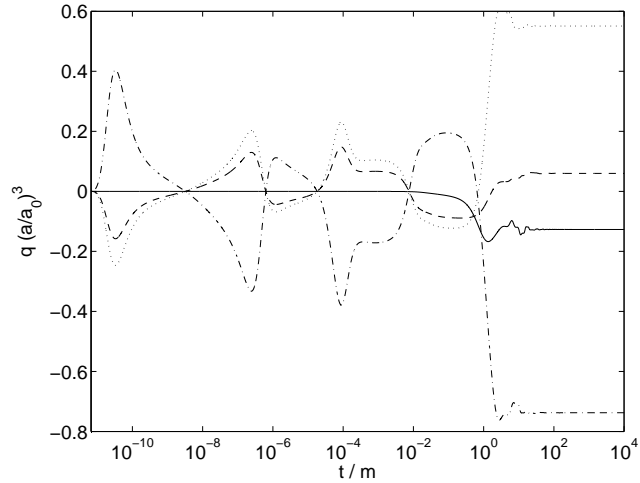}
\caption{\label{fig3} The charge densities in the co-moving volume for the
  three slepton fields with dashed line for the first, dash-dotted for the
  second and dotted for the third family. The total slepton charge density is
  given by the solid line. The parameters were $m_{\phi_i}^2=1$, $c_i=A=-1$,
  $a=i$ and $\lam_{ij}=\delta_{ij}$.}
\end{figure}

For simplicity, we have restricted our considerations to the case where
the Yukawa coupling matrix $\lam_{ij}$ is diagonal. We then find inumerical
solutions which are stable. In Figs. \ref{fig1},~\ref{fig2} we have
plotted the trajectories of the scalar fields $\phi_i$ in the complex plane.
In contrast with the one field case \cite{Dine:1995kz,Jokinen:2002xw}, the
scalar fields do not relax directly into the origin when $H\ll m_S$. This
is also implicit in Fig. \ref{fig3} where there is a non-zero charge for
the slepton fields right from the beginning. This feature can be ascribed
to a phase oscillation around the minimum of the potential.

In Fig. \ref{fig3} we have plotted the evolution of the lepton charge
densities in the co-moving volume of the slepton fields $L_i$ as parameterized
in \eq{struct325}. Note that the total lepton charge is in units of
$m_{\phi}^2 M_p$, where $m_{\phi}=100~\rm{GeV}$. The initial conditions
were again chosen by first solving the equations of motion during inflation.

\begin{figure}
\leavevmode
\centering
\vspace*{5,5cm}
\includegraphics{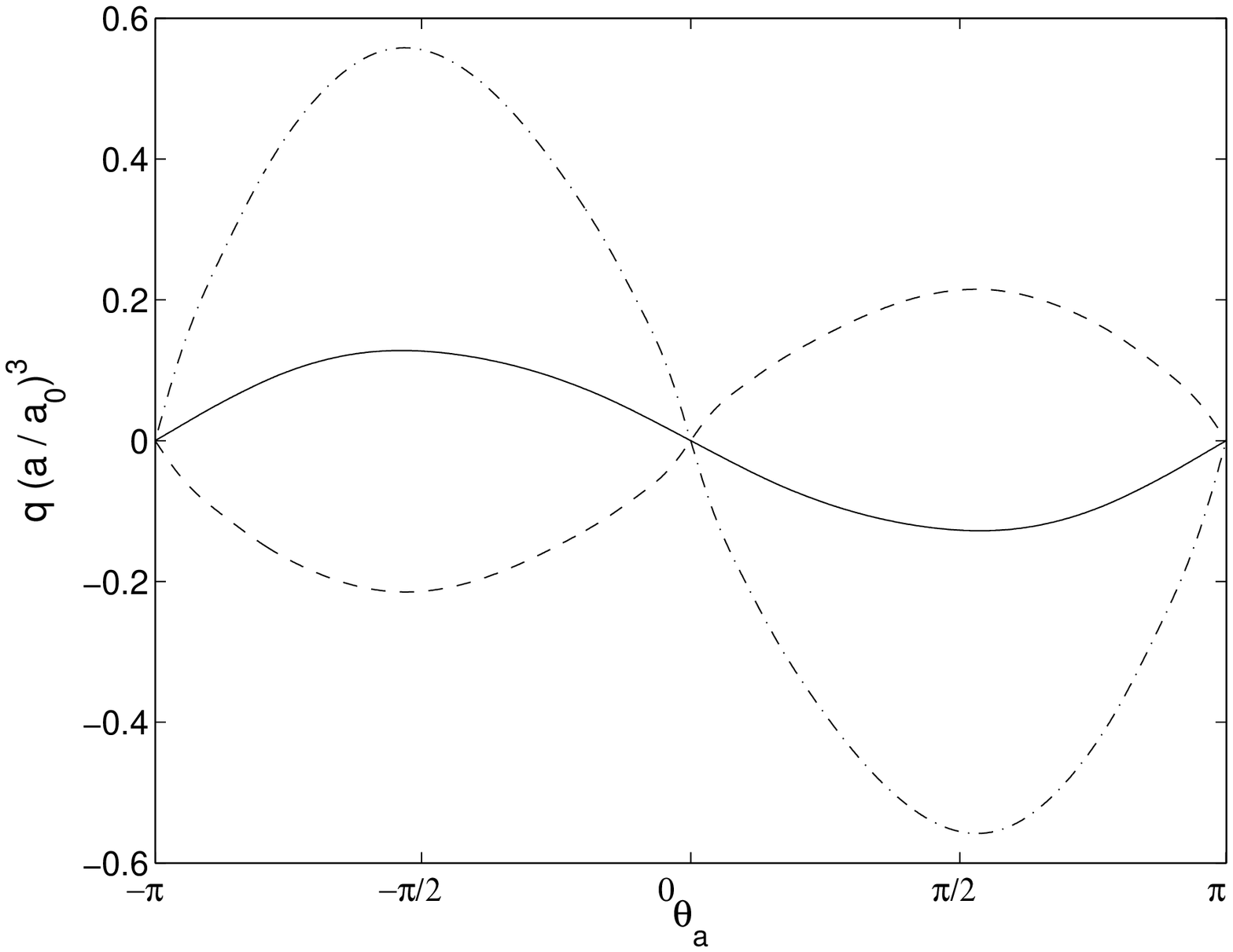}
\includegraphics{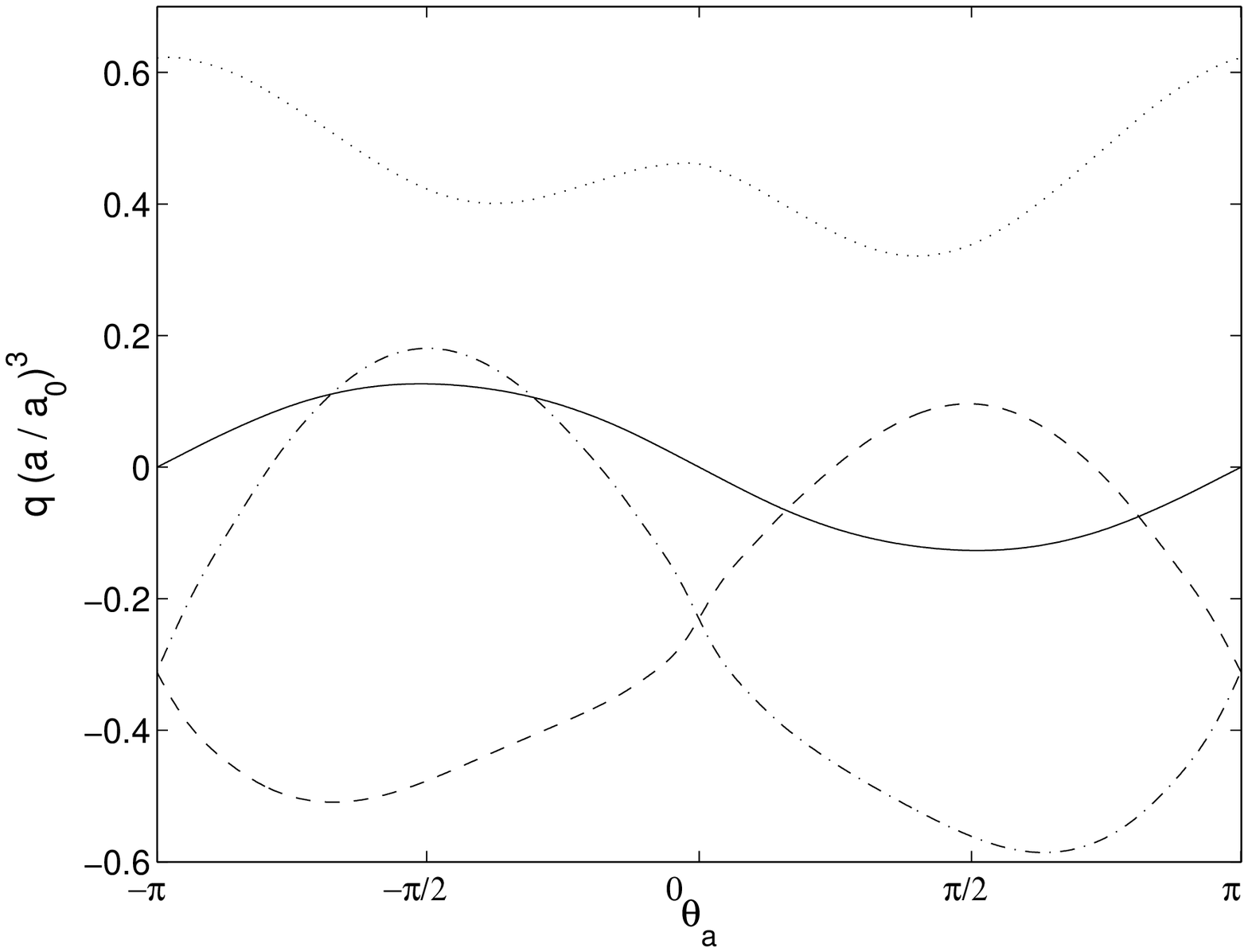}
\caption{\label{fig4} The charge density vs. the phase $\theta_a$ of $a$ with
  initial conditions during inflation $\phi_1=1$, $\phi_2=i$ and $\phi_3=-1$ in
  the upper figure and $\phi_1=1$, $\phi_2=i$ and $\phi_3=-1+i$ in the lower
  figure, where field values are in units of $(M_p H_I)^{1/2}$. The initial
  velocities vanish and the parameter values are $m_{\phi_i}^2=1$, $c_i=A=-1$,
  $|a|=1$ and $\lam_{ij}=\delta_{ij}$. Again solid line is for total charge,
  dashed line for $L_1$, dash-dot for $L_2$ and dotted for $L_3$. In the upper
  figure $L_1$ and $L_3$ are actually the same curve.}
\end{figure}

In Fig. \ref{fig4} we have plotted the dependence of charge density against
the phase $\theta_a$ of Hubble-induced A-term coefficient $a$ with two
different initial conditions during inflation. The different initial
conditions definitely affect the charge densities of the individual slepton
fields. However, although the initial conditions also affect the total charge
density, the effect is not as large as for the individual charges. If one
where to choose all the initial field values equal so that all the charge
densities would be equal, the total charge density evolution would differ from
those presented in  Fig. \ref{fig4} in magnitude but not in overall shape.

\begin{figure}
\leavevmode
\centering
\vspace{5,5cm}
\includegraphics{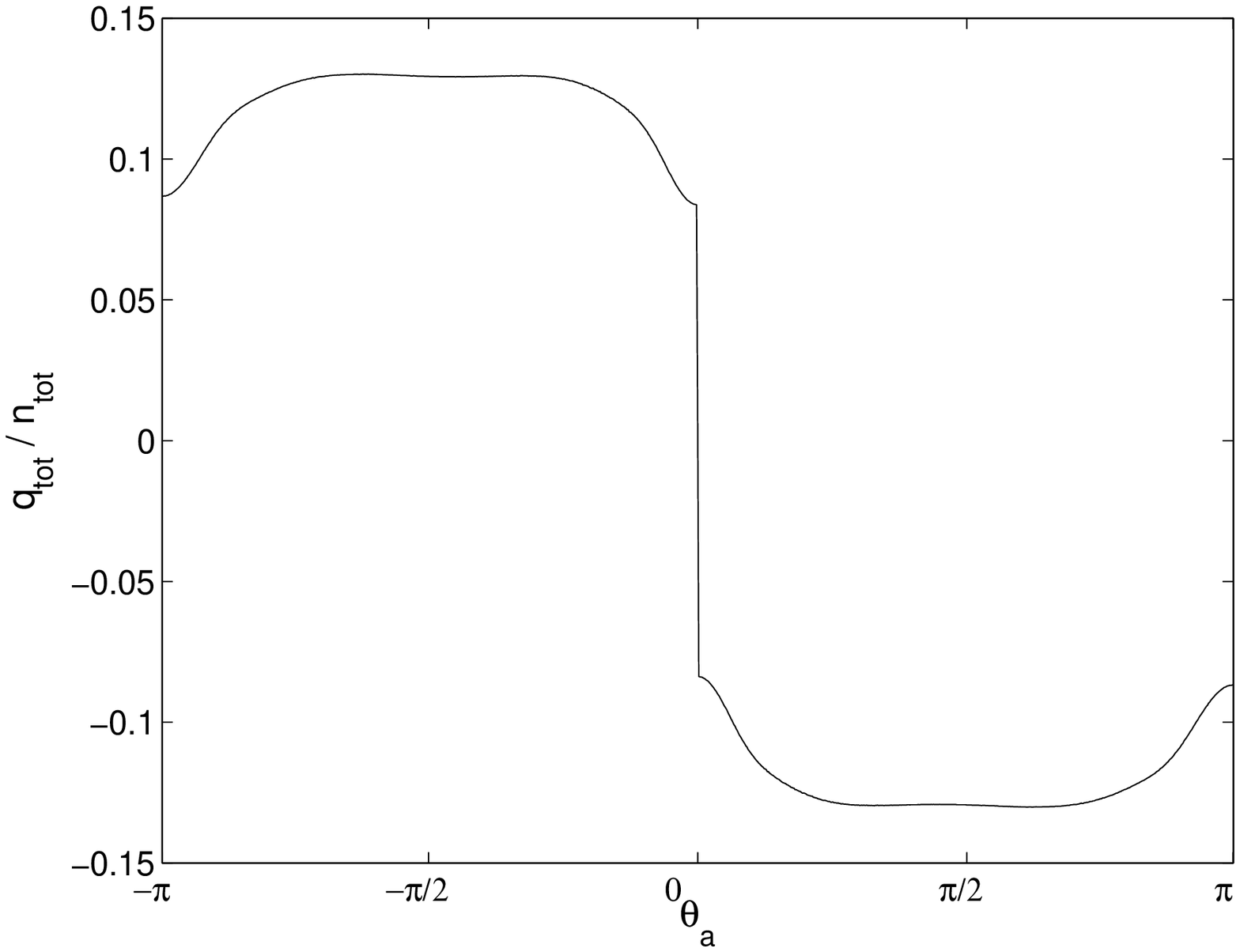}
\includegraphics{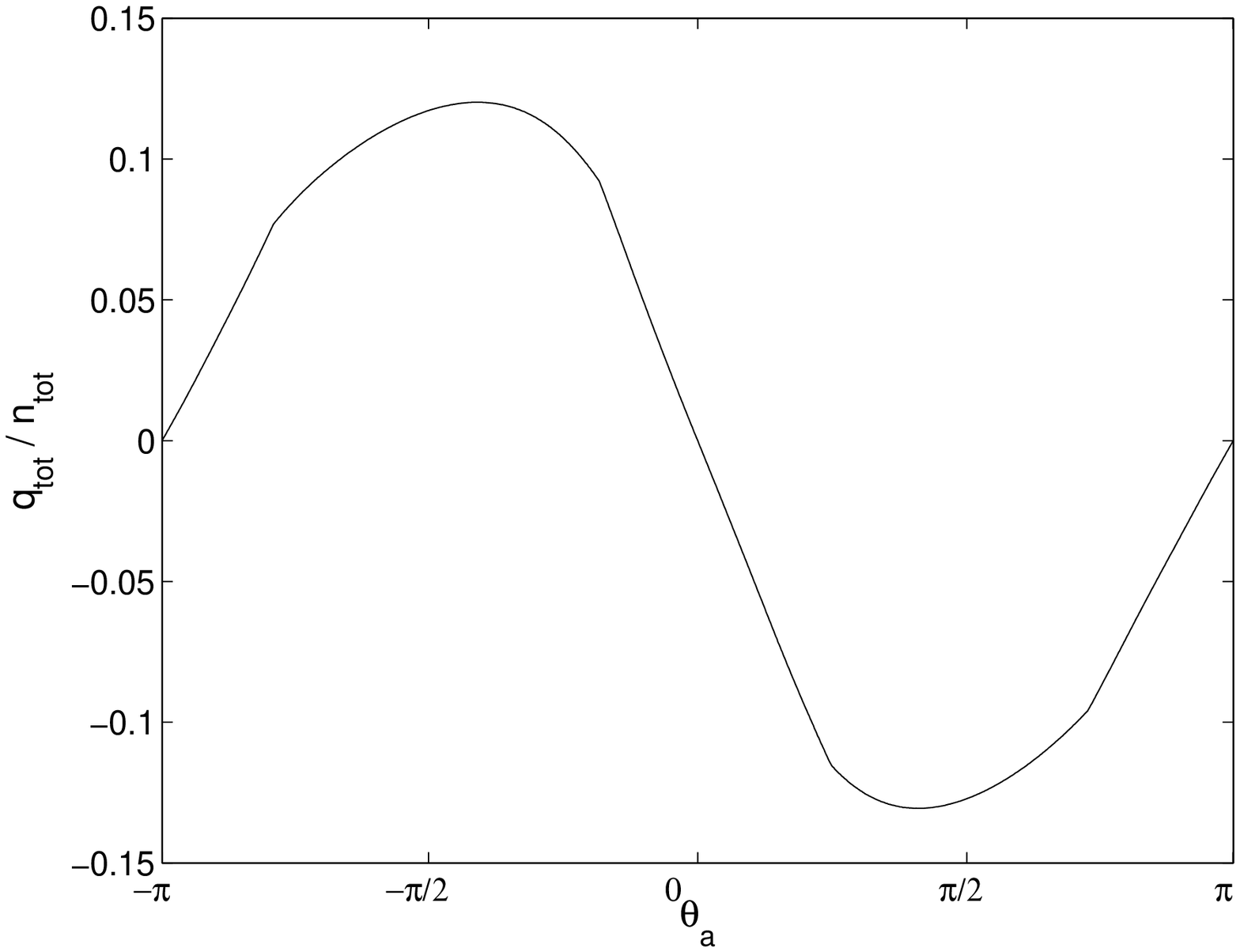}
\caption{\label{fig5} The charge asymmetries related to the cases of
  fig. \ref{fig3} with the same parameter values and initial conditions.}
\end{figure}

In Fig. \ref{fig5} we plot the charge asymmetry $q_{tot}/n_{tot}$, where
$q_{tot}=\sum q_i$ and $n_{tot}=\sum |q_i|$. One sees that for all these cases
one produces a  charge asymmetry of roughly $\sim 0.1$.

\section{Conclusions and Discussion}

The numerical studies indicate the produced baryon-to-entropy ratio is not
significantly changed when including all the lepton families to the $H_uL$
flat direction. It thus seems that the simple one-field case captures the
essential features of the Affleck-Dine mechanism, at least for $H_uL$.

However, there are some interesting details that do change. The first is that
the flat direction has non-minimal kinetic terms, which is due to the fact
that the sub-manifold spanned by the flat direction is curved. In the
one-field case non-minimality is absent because all the one-dimensional
manifolds are flat. Hence there is always a choice of coordinates where the
field metric is given by just the Kronecker delta. Another difference is that
there is phase motion of the fields already right after inflation. This is due
to the F-terms containing cross terms that break individual lepton
symmetries. However, the F-terms conserve the total lepton charge which is
produced only through the A-term. The individual charges may in general have
opposite signs and can give substantial contribution to the entropy, which is
proportional to the sum of absolute values of the individual charge
densities. The total charge is the sum of individual charge densities and thus
cancellations can occur.

Although the behaviour of a homogenous $H_uL$ background with multiple scalar
fields is not qualitatively different from the corresponding one-field case,
this may not be true if inhomogeneities are taken into account. For small
cosmological perturbations and multiple fields the gauge current condition of
\eq{grad323} becomes highly non-trivial. This can be seen as follows. $H_uL$
is described by three complex scalar fields or six real degrees of
freedom. The current condition \eq{grad323} defines a two-dimensional
submanifold in the six-dimensional space, so that there are four real
fluctuation degrees of freedom that would not obey \eq{grad323}. Hence there
would exist matter currents. This implies that the pure gauge field no longer
provides a solution for the equations of motion. As a consequence, there would
arise electric and magnetic gauge fields which could contribute to the CMB
perturbations. These might be of paramount importance for condensate
fragmentation, since the fragments (the ground states of which are Q-balls)
result from the non-linearities of the perturbations
\cite{Kusenko:1997zq,Enqvist:1997si}. The Q-ball properties
\cite{Coleman:1985ki} have been considered only for condensates with minimal
kinetic terms.

In this paper we focused on the $H_uL$ direction. For other directions more
complications may arise. For example, the $d=6$ flat direction
$(LL)(\bar{d}\bar{d}\bar{d})(\bar{u}\bar{d}\bar{d})$ has six different
superpotential terms that contribute to the lifting of the flat direction:
$d=4$ term $\bar{u}\bar{u}\bar{d}\bar{e}$, $d=5$ terms
$H_uL\bar{u}\bar{d}\bar{d}$ and $H_dL\bar{d}\bar{d}\bar{d}$ and $d=6$ terms
$\bar{u}\bar{d}\bar{d}QL\bar{d}$,
$(\bar{u}\bar{d}\bar{d})(\bar{u}\bar{d}\bar{d})$ and
$LL\bar{e}\bar{u}\bar{d}\bar{d}$. The $d=5,~6$ terms have $B-L=-2$ whereas the
$d=4$ term  has $B-L=0$. These will give rise to cross terms in the F term
which has a non-zero $B-L$ and therefore contributes to the total charge. Such
a situation occurs whenever the various superpotential terms with different
$B-L$ have at least one field in common.


\subsection*{Acknowledgements}
A.J. is supported by Magnus Ehrnrooth foundation.
K.E. is supported partly by the Academy of Finland grant no. 75065, and
A.M. is a CITA-National fellow. We thank Karsten Jedamzik for helpful
discussion.

\appendix

\section{Projection Matrices}

Projection of a vector $\mathbf{a}$ along a vector $\mathbf{b}$ is
given by
\be{proja1}
\mathbf{b}\,\frac{<\mathbf{b}|\mathbf{a}>}{|\mathbf{b}|^2}\,,
\ee
where $<\cdot|\cdot>$ is a (complex) scalar product. The operation of
projection corresponds to a matrix
\be{proja2}
P = \frac{\mathbf{b} \mathbf{b}^{\dagger}}{\mathbf{b}^{\dagger}\mathbf{b}}\,,
\ee
where we have marked the vector with a column matrix and $\dagger$ represents
the hermitian conjugation. Then \eq{proja1} is just $P\mathbf{a}$.

In the present paper the projection operators needed are somewhat
complicated since in principle they operate on the real basis of complex
vectors. Therefore we write complex scalar fields in terms of two real
fields $\phi_i = \psi_i + i\chi_i$. Then we can form a vector
\be{vectora3}
\left( \ba{ll} \psi \\ \chi \ea \right)\,,
\ee
which represents $\phi$ in the real basis. One can change the basis to
complex vectors $\phi$ and $\phi^*$ with
\be{vectora4}
\Phi \equiv \left( \ba{ll} \phi \\ \phi^* \ea \right) = \left( \ba{ll} 1 & i
  \\ 1 & -i \ea \right) \left( \ba{ll} \psi \\ \chi \ea \right)\,,
\ee
and construct a projection operator
\be{proja5}
P_1 = \frac{\Phi \Phi^{\dagger}}{\Phi^{\dagger} \Phi}\,.
\ee
We also need a projection operator along the vector orthogonal to $\Phi$
\be{proja6}
P_2 = \frac{\Psi \Psi^{\dagger}}{\Psi^{\dagger} \Psi}\,,
\ee
where
\be{vectora7}
\Psi = \left( \ba{ll} \phi \\ -\phi^* \ea \right)\,.
\ee
It is easy to check that $P_1^2 = P_1$, $P_2^2 = P_2$ and $P_1P_2 = P_2P_1 =
0$. With the help of these projection operators we can form the following
combinations
\bea{comba8}
\Phi^{\dagger}\Phi = \Psi^{\dagger}\Psi = 2 \sum_k |\phi_k|^2 \nn
\partial_{\mu} \Phi^{\dagger} P_1 \partial^{\mu}\Phi = \frac{\left[\sum_j
    (\phi_j^* \partial_{\mu} \phi_j + \phi_j \partial_{\mu}
    \phi_j^*)\right]^2} {2\sum_k |\phi_k|^2}\, \nn
\partial_{\mu} \Phi^{\dagger} P_2 \partial^{\mu}\Phi = \partial_{\mu}
\Psi^{\dagger} P_1 \partial^{\mu} \Psi = 
-\frac{\left[\sum_j (\phi_j^* \partial_{\mu} \phi_j - \phi_j \partial_{\mu}
    \phi_j^*)\right]^2} {2\sum_k |\phi_k|^2}\,.
\eea
With these identities we can write the Lagrangian as
\be{lagra9}
\mathcal{L} = \frac{1}{2}\partial_{\mu}\Phi^{\dagger}
\left(1+P_1-\frac{1}{2}P_2\right) \partial^{\mu}\Phi -V(\Phi,\Phi^{\dagger})\,.
\ee
The equations of motion read then
\bea{eqma10}
\left(1+P_1-\frac{1}{2}P_2\right) (\partial_{\mu}\partial^{\mu} \Phi +
3H\dot\Phi) + \dd{V}{\Phi^{\dagger}} - \partial_{\mu}\Psi
\frac{\Psi^{\dagger} \partial^{\mu} \Phi}{\Phi^{\dagger}\Phi} \nn +
\frac{\Psi}{\Phi^{\dagger}\Phi} \partial_{\mu}\Psi^{\dagger} P_2
\partial^{\mu} \Phi + \frac{\Phi}{\Phi^{\dagger}\Phi} \partial_{\mu}
\Phi^{\dagger} \left(1-P_1-\frac{1}{2}P_2\right) \partial^{\mu} \Phi = 0\,,
\eea
where $a$ is the scale factor of the FRW metric.

There is a formula for inverting a unit operator plus a sum of orthogonal
projection operators, given by
\be{proja11}
\left(1+\sum_i a_i P_i\right)^{-1} = 1 - \sum_i \frac{a_i}{1+a_i} P_i,
\ee
where $P_iP_j = \delta_{ij}P_i$ and $a_i \in \mathbb{C}\backslash\{-1\}$.
Using \eq{proja11} in \eq{eqma10}, we then obtain a form for the equation of
motion which is more easily comparable to the standard case:
\bea{eqma12}
\partial_{\mu}\partial^{\mu} \Phi + 3H\dot\Phi +
\left(1-\frac{1}{2}P_1+P_2\right) \dd{V}{\Phi^{\dagger}} -
\partial_{\mu}\Psi \frac{\Psi^{\dagger}
  \partial^{\mu}\Phi}{\Phi^{\dagger}\Phi} \nn + \frac{\Psi}{\Phi^{\dagger}\Phi}
\partial_{\mu}\Psi^{\dagger} P_2 \partial^{\mu} \Phi +
\frac{\Phi}{2\Phi^{\dagger}\Phi} \partial_{\mu} \Psi^{\dagger}
\left(1-P_1-\frac{3}{2}P_2\right)\partial^{\mu}\Phi = 0\,.
\eea
%




\section*{References}

\begin{thebibliography}{50}

\Bibitem{Buccella:1981ib}
F. Buccella, J. P. Derendinger, C. A. Savoy and S. Ferrara,
CERN-TH-3212
{\it Presented at 2nd Europhysics Study Conf. on the Unfication of the
  Fundamental Interactions, Erice, Italy, Oct 6-14, 1981};
F. Buccella, J. P. Derendinger, S. Ferrara and C. A. Savoy,
Phys.\ Lett.\ B {\bf 115} (1982) 375.

\Bibitem{gwr79}
M. Grisaru, W. Siegel, and M. Rocek, Nucl. Phys. B {\bf 159}, 429 (1979);
N. Seiberg, Phys. Lett. B {\bf 318}, 469 (1993).
[arXiv:hep-ph/9309335].

\Bibitem{Enqvist:2003gh}
K. Enqvist and A. Mazumdar,
Phys.\ Rept.\  {\bf 380}, 99 (2003)
[arXiv:hep-ph/0209244].

\Bibitem{Affleck:1984fy}
I.~Affleck and M.~Dine,
Nucl.\ Phys.\ B {\bf 249} (1985) 361.

\Bibitem{Dine:1995uk}
M. Dine, L. Randall and S. Thomas,
Phys.\ Rev.\ Lett.\  {\bf 75} (1995) 398
[arXiv:hep-ph/9503303].

\Bibitem{Dine:1995kz}
M. Dine, L. Randall and S. Thomas,
Nucl.\ Phys.\ B {\bf 458} (1996) 291
[arXiv:hep-ph/9507453].

\Bibitem{Kusenko:1997zq}
A. Kusenko,
Phys.\ Lett.\ B {\bf 405}, 108 (1997)
[arXiv:hep-ph/9704273];
A. Kusenko and M. E. Shaposhnikov,
Phys.\ Lett.\ B {\bf 418}, 46 (1998)
[arXiv:hep-ph/9709492].

\Bibitem{Enqvist:1997si}
K. Enqvist and J. McDonald,
Phys.\ Lett.\ B {\bf 425}, 309 (1998)
[arXiv:hep-ph/9711514];
K. Enqvist and J. McDonald,
Nucl.\ Phys.\ B {\bf 538}, 321 (1999)
[arXiv:hep-ph/9803380].

\Bibitem{Jokinen:2002xw}
A. Jokinen,
arXiv:hep-ph/0204086.

\Bibitem{Kasuya:1999wu}
S.~Kasuya and M.~Kawasaki,
Phys.\ Rev.\ D {\bf 61} (2000) 041301
[arXiv:hep-ph/9909509];
S.~Kasuya and M.~Kawasaki,
Phys.\ Rev.\ D {\bf 62} (2000) 023512
[arXiv:hep-ph/0002285];
K.~Enqvist, A.~Jokinen and J.~McDonald,
Phys.\ Lett.\ B {\bf 483} (2000) 191
[arXiv:hep-ph/0004050];
K. Enqvist, A. Jokinen, T. Multamaki and I. Vilja,
Phys.\ Rev.\ D {\bf 63}, 083501 (2001)
[arXiv:hep-ph/0011134];
S.~Kasuya and M.~Kawasaki,
Phys.\ Rev.\ D {\bf 64} (2001) 123515
[arXiv:hep-ph/0106119];
T.~Multamaki and I.~Vilja,
Phys.\ Lett.\ B {\bf 535} (2002) 170
[arXiv:hep-ph/0203195].

\Bibitem{Enqvist:2002rj}
K. Enqvist, S. Kasuya and A. Mazumdar,
Phys.\ Rev.\ Lett.\  {\bf 89}, 091301 (2002)
[arXiv:hep-ph/0204270].
K. Enqvist, S. Kasuya and A. Mazumdar,
Phys.\ Rev.\ D {\bf 66}, 043505 (2002)
[arXiv:hep-ph/0206272].

\Bibitem{Enqvist:1998pf}
K. Enqvist and J. McDonald,
Phys.\ Rev.\ Lett.\  {\bf 83}, 2510 (1999)
[arXiv:hep-ph/9811412].
K. Enqvist and J. McDonald,
Phys.\ Rev.\ D {\bf 62}, 043502 (2000)
[arXiv:hep-ph/9912478].
M. Kawasaki and F. Takahashi,
Phys.\ Lett.\ B {\bf 516}, 388 (2001)
[arXiv:hep-ph/0105134].

\Bibitem{Enqvist:2002rf}
K. Enqvist, S. Kasuya and A. Mazumdar,
Phys.\ Rev.\ Lett.\  {\bf 90}, 091302 (2003)
[arXiv:hep-ph/0211147];
M.~Postma,
Phys.\ Rev.\ D {\bf 67}, 063518 (2003)
[arXiv:hep-ph/0212005];
K. Enqvist, A. Jokinen, S. Kasuya and A. Mazumdar,
arXiv:hep-ph/0303165;
S.~Kasuya, M.~Kawasaki and F.~Takahashi,
arXiv:hep-ph/0305134;
M.~Postma and A.~Mazumdar,
arXiv:hep-ph/0304246;
K.~Hamaguchi, M.~Kawasaki, T.~Moroi and F.~Takahashi,
arXiv:hep-ph/0308174;
K.~Enqvist, S.~Kasuya and A.~Mazumdar,
arXiv:hep-ph/0311224.

\bibitem{Enqvist:2003uk}
K.~Enqvist, A.~Mazumdar and M.~Postma,
Phys.\ Rev.\ D {\bf 67}, 121303 (2003)
[arXiv:astro-ph/0304187];
A.~Mazumdar and M.~Postma,
Phys.\ Lett.\ B {\bf 573}, 5 (2003)
[arXiv:astro-ph/0306509].



\Bibitem{Gherghetta:1995dv}
T. Gherghetta, C. F. Kolda and S. P. Martin,
model,''
Nucl.\ Phys.\ B {\bf 468} (1996) 37
[arXiv:hep-ph/9510370].

\Bibitem{Senami:2001qn}
M.~Senami and K.~Yamamoto,
Phys.\ Lett.\ B {\bf 524} (2002) 332
[arXiv:hep-ph/0105054];
M.~Senami and K.~Yamamoto,
Phys.\ Rev.\ D {\bf 66} (2002) 035006
[arXiv:hep-ph/0205041];
M.~Senami and K.~Yamamoto,
Phys.\ Rev.\ D {\bf 67} (2003) 095005
[arXiv:hep-ph/0210073];
M.~Senami and K.~Yamamoto,
arXiv:hep-ph/0305202;

\Bibitem{Nilles:1983ge}
H. P. Nilles,
Phys.\ Rept.\  {\bf 110} (1984) 1.

\Bibitem{Sartori:1981zj}
G. Sartori,
J.\ Math.\ Phys.\  {\bf 24} (1983) 765;
R. Gatto and G. Sartori,
Phys.\ Lett.\ B {\bf 118} (1982) 79;
C. Procesi and G. W. Schwarz,
Phys.\ Lett.\ B {\bf 161} (1985) 117;
R. Gatto and G. Sartori,
Phys.\ Lett.\ B {\bf 157} (1985) 389;
R. Gatto and G. Sartori,
Commun.\ Math.\ Phys.\  {\bf 109} (1987) 327;
G. Sartori,
Mod.\ Phys.\ Lett.\ A {\bf 4} (1989) 91.

\Bibitem{Brax:2001an}
P. Brax and C. A. Savoy,
arXiv:hep-th/0104077.

\Bibitem{Kolda:1998kc}
C. F. Kolda and J. March-Russell,
Phys.\ Rev.\ D {\bf 60} (1999) 023504
[arXiv:hep-ph/9802358].

\bibitem{Coleman:1985ki}
S.~R.~Coleman,
Nucl.\ Phys.\ B {\bf 262} (1985) 263
[Erratum-ibid.\ B {\bf 269} (1986) 744].

\end{thebibliography}
\end{document}